\documentstyle[amssymb,pra,aps]{revtex}
\begin{document}
\draft
\title{{{ Quantum to Classical Transition from the Cosmic Background Radiation }}}
\author{M.C. de Oliveira$^1$, N.G. Almeida$^2$, S.S. Mizrahi$^2$, and M.H.Y. Moussa$^2$}
\address{
$^1$Department of Physics, University of Queensland,QLD 4072, Brisbane,
Australia.\\
$^2$Departamento de F\'{\i }sica, Universidade Federal de S\~{a}o
Carlos\\
Via Washington Luis, km 235, S\~{a}o Carlos 13565-905, SP, Brasil}
\date{\today}
\maketitle
\begin{abstract}
{We have revisited the Ghirardi-Rimini-Weber-Pearle (GRWP) approach for
continuous dynamical evolution of the state vector for a macroscopic object.
Our main concern is to recover the decoupling of the state vector dynamics
for the center-of-mass (CM) and internal motion, as in the GRWP model, but
within the framework of the standard cosmology. In this connection we have
taken the opposite direction of the GRWP argument, that the cosmic
background radiation (CBR) has originated from a fundamental stochastic
hitting process. We assume the CBR as a clue of the Big Bang, playing a main
role in the decoupling of the state vector dynamics of the CM and internal
motion. In our model, instead of describing a continuous spontaneous
localization (CSL) of a system of massive particles as proposed by Ghirardi,
Pearle and Rimini, the It\^{o} stochastic equation accounts for the
intervention of the CBR on the system of particles. Essentially, this
approach leads to a pre-master equation for both the CBR and particles
degrees of freedom. The violation of the principle of energy conservation
characteristic of the CSL model is avoided as well as the additional
assumption on the size of the GRWP's localization width necessary to reach
the decoupling between the collective and internal motions. Moreover,
realistic estimation for the decoherence time, exhibiting an interesting
dependence on the CBR temperature, is obtained. From the formula for the
decoherence time it is possible to analyze the transition from micro to
macro dynamics in both the early hot Universe and the nowadays cold one. The
entropy of the system under decoherence is analyzed and the emergent
`pointer basis' is discussed. In spite of not having imposed a privileged
basis, in our model the position still emerges as the preferred observable
as in the CSL model.}
\end{abstract}


%
%% BEGIN ABSTRACT %%%%%%%%%%%%%%%%%%%%%%%%%%%%%%%%%%%%%%%%%%%%%%%%%%
%

%
%%%%%%%%%%%%%%%%%%%%%%%%%%%%%%%%%%%%%%%%%%%%%%%%%%%%%%%%%%%%%%%%%%
%

\section{Introduction}

%
%%%%%%%%%%%%%%%%%%%%%%%%%%%%%%%%%%%%%%%%%%%%%%%%%%%%%%%%%%%%%%%%%%
In the last decade several proposals to modify the standard Hamiltonian
dynamics, ranging from master equations to stochastic quantum mechanics,
have been advanced to try to set up an unified description for microscopic
and macroscopic physical phenomena. In the pioneer work by Ghirardi, Rimini,
and Weber \cite{grw}, {\it quantum mechanics with spontaneous localization}
(QMSL), the state vector collapse, leading from quantum to classical
dynamics results from the instantaneous action of a spontaneous random
hitting process. Such a Poisson process is described by a ``localization''
operator, a gaussian function acting on each microscopic constituent of any
system. The localization operator carries two free parameters; a mean
frequency $\lambda $ and a localization width $\alpha ^{-1/2}$, understood
as new constants of nature (the {\it spontaneous localization} is argued to
be a fundamental physical process). Through these basic assumptions the QMSL
consists in an explicit model allowing an unified description for
microscopic and macroscopic systems. It forbids the occurrence of linear
superposition of states localized in far away spatial regions and induces a
dynamics that agree with the predictions of classical mechanics.

Pursuing the program of the QMSL model, Diosi \cite{diosi} presented an
interesting connection between the original GRW hitting process and a
modified Schr\"{o}dinger equation. Another significant achievement
concerning a dynamical reduction model, a stochastic equation for physical
ensemble, was reported by Gisin \cite{gisin}. Next, Pearle \cite{pearle}
described the QMSL model through an It\^{o} stochastic differential
equation. Basically, Pearle replaced the Poisson process of instantaneous
hits in GRW model by a Markov process described as a stochastic modification
of the Schr\"{o}dinger equation, so that a continuous evolution of the state
vector was accomplished. By considering a specific choice of the operators
defining the Markov process (expressed in terms of creation and annihilation
operators), Ghirardi, Pearle, and Rimini \cite{gpr} have described the
mechanism known as {\it continuous spontaneous localization} (CSL) of
systems of identical particles (the QMSL model has consistency only in the
case of systems of distinguishable particles).

Other investigations dealing with dynamical reduction models have recently
been considered \cite{grigolini}, among them it is worth to mention the
model for {\it intrinsic} decoherence proposed by Milburn \cite{milburn}.
While in the GRWP model the addition of stochastic terms in the
Schr\"{o}dinger evolution automatically destroys the quantum coherence of
the physical properties of the system that attain a macroscopic level, the
modification of the Liouville equation proposed by Milburn destroys the
coherence even at microscopic level.

In the CSL model the It\^{o} stochastic equation for the evolution of the
state vector reads
\begin{equation}
d|\psi \rangle =\left( -\frac i\hbar Hdt+dh-\frac 12\overline{(dh)^2}\right)
|\psi \rangle ,  \label{1}
\end{equation}
where $dh$ is a linear self-adjoint operator, whose random fluctuation may
increase or decrease the norm of the state vector. Using It\^{o} formula
(with the notation $|d\psi \rangle \equiv d|\psi \rangle $),
\begin{equation}
d\parallel \psi \parallel ^2=\langle \psi |d\psi \rangle +\langle d\psi
|\psi \rangle +\overline{\langle d\psi |d\psi \rangle },  \label{2}
\end{equation}
it is easy to see that Eq. (\ref{1}) does not conserve the norm of $|\psi
\rangle $. Thus, the introduction of a norm conserving nonlinear process is
mandatory. This process, whose random operator depends on the state vector,
reads
\begin{equation}
d|\phi \rangle =\left( -\frac i\hbar Hdt+dh_\phi -\frac 12\overline{(dh_\phi
)^2}\right) |\phi \rangle .  \label{3}
\end{equation}
Now, it is necessary to distinguish between {\it raw} ( Eq. (\ref{1})) and
{\it physical} ( Eq. (\ref{3})) ensembles of state vectors to correctly
understand the effect of the non-Hamiltonian terms. To this end a precept is
adopted, namely, that the square norm of each (unnormalized) state vector
represents the weight associated with that (normalized) state vector in the
ensemble coming from the It\^{o} stochastic equation \cite{pearle,gpr}. This
precept is a generalization of the GRW assumption that the frequency of hits
is proportional to the squared norm of the state vector. Therefore, in the
GRW prescription the quantum theory prediction about the associated
probabilities in a measurement process is recovered. By considering such a
precept for the physical ensemble, the linearity of the {\it raw} equation
and the Markov nature of the It\^{o} stochastic process leads to the {\it %
physical} stochastic differential equation for the $N$-particle state vector
\begin{equation}
d|\Psi _N\rangle =\left( -\frac i\hbar Hdt+{\bf Z}.d{\bf B}-\frac 12\gamma
{\bf Z}^{\dagger }.{\bf Z}dt\right) |\Psi _N\rangle ,  \label{4}
\end{equation}
where ${\bf Z}\equiv \left\{ Z_i\right\} $ are operators on the Hilbert
space of the system and the set of random operators ${\bf B}\equiv \left\{
B_i\right\} $ is characterized through a real Wiener process, satisfying the
following means and correlations over ensemble
\begin{equation}
\overline{dB_i}=0,~~\overline{dB_idB_j}=\gamma \delta _{ij}dt.  \label{5}
\end{equation}
The statistical operator $\rho _N=\overline{|\Psi _N\rangle \langle \Psi _N|}
$ of the {\it physical} ensemble and its evolution equation are directly
obtained from Eq. (\ref{4}); using the It\^{o} calculus in evaluating $d\rho
_N/dt$ one gets
\begin{equation}
\frac{d\rho _N}{dt}=-\frac i\hbar \left[ H,\rho _N\right] +\gamma {\bf Z}%
\rho _N.{\bf Z}^{\dagger }-\frac \gamma 2\left\{ {\bf Z}^{\dagger }.{\bf Z}%
,\rho _N\right\} ,  \label{6}
\end{equation}
which is exactly the Lindblad \cite{lindblad} form for the generator of a
quantum dynamical semigroup.

In the present work our main concern is to achieve {\em the decoupling
between the state vector dynamics of the center-of-mass (CM) and internal
motion of a system of particles}. In the GRWP model this {\em decoupling}
results from a hypothesis of spontaneous localization of the system's wave
function due to a fundamental stochastic hitting process on the particles,
which induces an {\em increase of total mean energy of the Universe} claimed
to be the origin of the Cosmic Background Radiation (CBR). Contrarily to
this argument, in the present work we assume the point of view of standard
cosmology: the nowadays CBR is a clue that the Universe began its expansion
from a Big Bang \cite{turner}. This assumption is introduced with the
purpose to avoid the unconventional increase of the total mean energy of the
Universe. Formally, we hypothesize that the state vector, the Hamiltonian $H$
and operators ${\bf Z}$, ${\bf Z}^{\dagger }$ in Eq.(\ref{4}) represent
both, the system of particle and CBR radiation; the set of random functions $%
\{B_i\}$ describes the intervention of the CBR on the system and substitute
the spontaneous localization process. Instead of elaborating on the formal
microscopic problem of the interaction of a system with a reservoir \cite
{caldeira}, we assume {\it ad-hoc} that the evolution of the system of
particles, under the influence of the CBR, is described by an It\^{o}
equation having stochastic coupling parameters.

Therefore, in the present {\it conservative} continuous reduction model (the
total energy of system plus CBR is conserved) we argue that: 1) the increase
or decrease of the system's mean energy is attributed to the CBR; 2) the
positional space is not privileged with respect to the momentum space, as
required when the localization operator is involved; 3) we do not claim for
an {\it additional assumption} to decouple the collective and internal
motion, namely the width parameter $\alpha ^{-1/2}\thicksim 10^{-5}cm$ in
the CSL model; 4) as above-mentioned, more admissible results are obtained
for decoherence times, while in the CSL model the value $10^{-7}s$ obtained
for a system of particles to undergo from quantum to classical dynamics
seems to be too large (as well as the localization width $\alpha
^{-1/2}\thicksim 10^{-5}cm$ also seems too large when considering typical
atomic distances about $10^{-8}cm$, or even superposition of the
center-of-mass coordinate different by more than about $\alpha ^{-1/2}$ \cite
{miled}), and finally, 5) instead of the two free parameters required in the
GRWP model ($\alpha ^{-1/2}$ and the mean frequency $\lambda $), the random
function describing the interaction between the system and the CBR carries
just a single strength parameter with dimension of inverse of time. In fact,
the coupling constant of the CBR photons to the $N$-particle system, as the
strength parameter in the GRWP model, defines the inverse of a
characteristic time which is associated to the net effect of the random
pseudo-``potential'' $dh$ \cite{tony}. Also, as in the GRWP model, our
strength parameter is such small that nothing changes in the Hamiltonian
dynamics of a single particle even in the case in which it has an extended
wave function \cite{gpr}.

Finally, we mention that Joos and Zeh \cite{joos}, have previously argued
that scattering of photons even at a relatively low temperature can induce
the localization of the wave packet of a macroscopic system. So, their
treatment, based on a master equation proposed by Wigner \cite{wigner},
suggests that the intergalactic cold CBR cannot simply be neglected \cite
{kiefer}. The model here presented goes exactly on this point, {\it i.e.},
we consider the process of random scattering of the CBR photons by a system
of particles as responsible for the superselection rules and the micro to
macro transition of its dynamical description. In this way, despite inducing
the superselection rules the CBR also induces the mechanism of separating
the center-of-mass (CM) coordinate from the internal motion. Besides, we
present a brief cosmological analysis of our results, discussing the roles
played by both the CBR temperature and the number of particles of the
system, in its way from quantum to classical dynamics, as the universe
evolved from a hot to a cold state.

In Section II we briefly review the GRWP model presenting its main
achievements. In Section III we construct our model: beginning from an
It\^{o} stochastic equation we derive a pre-master equation for a system of $%
N$ particles and the CBR; tracing over the CBR degrees of freedom we obtain
a master equation for the system of particles only and In Section IV we show
that structurally it shows exactly the Lindblad form. In Section V we
estimate the coupling parameter and in Section VI we estimate the
decoherence time for the system of particles. In Section VII we show that at
low temperature limit our master equation and the GRWP It\^{o} equation are
equivalent, thus this last one is a particular situation of the former;
these equations allow the decoupling of the state vector dynamics into two
separate equations, one for the CM and the other for the internal motion. In
section VIII we calculate the entropy and analyze the problem of selection
of a preferred basis. Finally, in Section IX we present a summary and
conclusions. %
%%%%%%%%%%%%%%%%%%%%%%%%%%%%%%%%%%%%%%%%%%%%%%%%%%%%%%%%%%%%%%%%%%%%%%%%%%%%%%%%%%%

\section{The Ghirardi-Rimini-Weber-Pearle Model}

%%%%%%%%%%%%%%%%%%%%%%%%%%%%%%%%%%%%%%%%%%%%%%%%%%%%%%%%%%%%%%%%%%%%%%%%%%%%%%%%%%%
%
As explained in the introduction, in CSL model the random operator $dh$
contains in its definition the length parameter $\alpha ^{-1/2}$ and a
strength parameter $\zeta $ which is related to the mean hitting frequency $%
\lambda $. In this section we present a brief review of the CSL model as a
class of Markov processes in Hilbert space \cite{gpr}. We will consider a
system of $N$ identical particles so that the localization operator must
involve globally the whole set of particles in order to preserve the
symmetry properties of the wave function \cite{web}. For this purpose let us
consider the creation and annihilation field operators $a^{\dagger }({\bf q}%
,s)$, $a({\bf q},s)$ of a particle at the point ${\bf q}$ in some reference
frame with spin component $s$, satisfying the canonical commutation or
anticommutation relations. From these operators a locally averaged number
density operator is defined as
\begin{equation}
N({\bf x})=(\frac \alpha {2\pi })^{3/2}\sum_s\int d^3{\bf q}\exp \left[ -%
\frac 12\alpha ({\bf q}-{\bf x})^2\right] a^{\dagger }({\bf q},s)a({\bf q}%
,s).  \label{7}
\end{equation}
The operator $N({\bf x})$ is self adjoint and its commutator for different
values of ${\bf x}$ vanishes. The total number operator is defined as $%
N=\int d^3{\bf x}N({\bf x})$, and the symmetrized (antisymmetrized) states
containing $n$ particles at the indicated positions,
\begin{equation}
\mid {\bf q},s\rangle ={\cal N}a^{\dagger }({\bf q}_1,s_1)...a^{\dagger }(%
{\bf q}_n,s_n)\mid 0\rangle ,  \label{8}
\end{equation}
constitutes the normalized common eigenvectors related to the eigenvalue
equation $N({\bf x})\mid {\bf q},s\rangle =n_{{\bf x}}\mid {\bf q},s\rangle $%
, with
\begin{equation}
n_{{\bf x}}=(\frac \alpha {2\pi })^{3/2}\sum_{i=1}^N\exp \left[ -\frac 12%
\alpha ({\bf x}-{\bf q}_i)^2\right] .  \label{9}
\end{equation}

Applying \cite{gpr,web} the stochastic process established by Eq. (\ref{4})
to a system of identical particles and considering the locally averaged
density operator defined by Eq. (\ref{7}), one gets the physical stochastic
nonlinear differential equation for the state vector as
\begin{equation}
d\mid \psi _N\rangle =\left[ -iHdt+\int d^3{\bf x}N({\bf x})dB({\bf x})-%
\frac 12\zeta \int d^3{\bf x}N^2({\bf x})dt\right] \mid \psi _N\rangle .
\label{10}
\end{equation}
where the Wiener process $B({\bf x})$ satisfies
\begin{mathletters}
\label{11}
\begin{eqnarray}
\overline{dB({\bf x})} &=&0,\   \label{11a} \\
\overline{dB({\bf x})dB({\bf y})} &=&\zeta \delta ^3({\bf x}-{\bf y})dt.
\label{11b}
\end{eqnarray}
From Eq. (\ref{10}) the evolution equation of the $N$-particle statistical
operator obtained from It\^{o} calculus reads
\end{mathletters}
\begin{equation}
\frac{\partial \rho _N}{\partial t}=-i[H,\rho _N]+\zeta \int d^3{\bf x}N(%
{\bf x})\rho _NN({\bf x})-\frac 12\zeta \left\{ \int d^3{\bf x}N^2({\bf x}%
),\rho _N\right\} .  \label{12}
\end{equation}
and it can be checked that taking $\lambda =\zeta (\alpha /4\pi )^{3/2}$,
Eq. (\ref{12}) reduces to the correspondent equation for a single particle
considered in the QMSL model.

To discuss the physical implications of the modified dynamical equation (\ref
{10}), the separation of the CM motion will be made. If ${\bf Q}$ is the CM
coordinate of the system and $\widetilde{{\bf q}}_i$ its internal
coordinates (measured from the CM of the particles), one can define the
particle coordinates as
\begin{equation}
{\bf q}_i={\bf Q}+\widetilde{{\bf q}}_i(\left\{ {\bf r_i}\right\} ),
\label{13}
\end{equation}
where $\left\{ {\bf r_i}\right\} $ represents a set of $3N-3$ independent
variables. In the GRWP model the set $\left\{ {\bf r_i}\right\} $ does not
contain macroscopic variables. As a consequence, assuming that the
Hamiltonian can be written as $H=H_Q+H_{r_i}$, we consider the wave function
\begin{mathletters}
\label{14}
\begin{eqnarray}
\phi ({\bf q},s) &=&\Psi ({\bf Q})\chi ({\bf r_i},s),  \label{14a} \\
\chi ({\bf r_i},s) &=&{{%
%TCIMACRO{\binom{A }{B}}
%BeginExpansion
{A  \choose B}%
%EndExpansion
}}\Delta ({\bf r_i},s),  \label{14b}
\end{eqnarray}
where the symbol ${{%
%TCIMACRO{\binom{A }{B}}
%BeginExpansion
{A  \choose B}%
%EndExpansion
} }$ specify the symmetrization or antisymmetrization of the internal
coordinate wave function. Under the assumption that the length parameter $%
\alpha ^{-1/2}$ is such that the internal wave function $\Delta ({\bf r_i},s)
$ is sharply peaked around the value ${\bf r_i}_0$ of ${\bf r}$ (with
respect to $\alpha ^{-1/2}$ ), the action of the operator $N({\bf x})$ on
the wave function (\ref{14a}) turns out to be
\end{mathletters}
\begin{equation}
N({\bf x})\Psi ({\bf Q})\chi ({\bf r_i},s)=F({\bf Q}-{\bf x})\Psi ({\bf Q}%
)\chi ({\bf r_i},s),  \label{15}
\end{equation}
with
\begin{equation}
F({\bf Q}-{\bf x})=\sum_i(\frac \alpha {2\pi })^{3/2}\exp \left\{ -\frac 12%
\alpha \left[ {\bf Q}+\widetilde{{\bf q}}_i({\bf r}_0)-{\bf x}\right]
^2\right\} .  \label{16}
\end{equation}
Therefore, the operator $N({\bf x})$ acts only on $\Psi $ and the separately
normalized wave functions $\Psi $ and $\chi $ satisfy the equations
\begin{mathletters}
\label{17}
\begin{eqnarray}
d\Psi &=&\left[ -iH_Qdt+\int d^3{\bf x}F({\bf Q}-{\bf x})dB({\bf x})-\frac 12%
\zeta \int d^3{\bf x}F^2({\bf Q}-{\bf x})dt\right] \Psi ,  \label{17a} \\
d\chi &=&-iH_{r_i}\chi dt.  \label{17b}
\end{eqnarray}

By assuming a large enough length parameter and an internal wave function
which is independent of the macroscopic variables, the internal motion
decouples as in the absence of the stochastic terms in Eq. (\ref{10}). From
this fact, the reduction rates which are characteristic of the GRWP theory
together with the position and momentum spreading can be obtained. In
particular, in the positional representation of Eq. (\ref{12}), it is
possible to verify with the help of the macroscopic density approximation
and the sharp scanning approximation \cite{gpr}, that the macroscopic
frequency associated to the system of identical particles is
\end{mathletters}
\begin{equation}
\Gamma =\zeta D_0n_{out}.  \label{18}
\end{equation}
Here a homogeneous macroscopic body of density $D_0$ was considered and $%
n_{out}$ is the number of particles of the body at position ${\bf Q^{\prime }%
}$ that do not lie in the volume occupied by the body at position ${\bf %
Q^{\prime \prime }}$. In the case of distinguishable particles, one gets the
direct result
\begin{equation}
\lambda _{CM}=n\lambda ,  \label{19}
\end{equation}
$n$ being the total number of particles, so that for a typical macroscopic
number $n\thickapprox 10^{23}$, one obtains $\lambda _{CM}\thickapprox
10^{-7}s$, as mentioned above.

The position and momentum spreading obtained from the approximations leading
to Eq. (\ref{18}), are written as
\begin{mathletters}
\label{20}
\begin{eqnarray}
\langle Q_i^2\rangle &=&\langle Q_i^2\rangle _s+\zeta \delta _i\frac{\hbar ^2%
}{6M^2}t^3,  \label{20a} \\
\langle P_i^2\rangle &=&\langle P_i^2\rangle _s+\frac 12\zeta \delta _i\hbar
^2t,  \label{20b}
\end{eqnarray}
where the suffix $s$ indicates the Schr\"{o}dinger evolution, and
\end{mathletters}
\begin{equation}
\delta _i=\int d^3{\bf y}\left( \frac{\partial F({\bf y})}{\partial y_i}%
\right) ^2.  \label{21}
\end{equation}

Now, using the macroscopic density approximation applied to the identical
particles system, Eq. (\ref{16}) is modified to
\begin{equation}
F({\bf Q}-{\bf x})=\int d^3\widetilde{{\bf y}}D(\widetilde{{\bf y}})(\frac %
\alpha {2\pi })^{3/2}\exp \left[ -\frac 12\alpha ({\bf Q}+\widetilde{{\bf y}}%
-{\bf x})^2\right] ,  \label{22}
\end{equation}
where $D({\bf y})$ is the number of particles per unit volume in the
neighborhood of the point ${\bf y}={\bf Q}+\widetilde{{\bf y}}$. The
evaluation of the factor $\delta _i$ for the case of a homogeneous
macroscopic box containing the $N$ particles, through the Eq. (\ref{22})
gives the result \cite{gpr}
\begin{equation}
\delta _i=(\alpha /\pi )^{1/2}D_0^2S_i,  \label{23}
\end{equation}
where $S_i$ is the transversal section of the macroscopic box.

From Eq. (\ref{20b}) it is evident that the momentum variance implies that
the CM energy increases per unit time as
\begin{equation}
\frac{\Delta E}t=\frac{\zeta \delta _i\hbar ^2}M\sim
10^{-32}(g~cm~s^{-1})S_i~cm^{-2},  \label{24}
\end{equation}
with the GRWP choice $\alpha ^{-1/2}\sim 10^{-5}~cm$ together with $D_0\sim
10^{24}~cm^{-3}$. From the requirement that the macroscopic frequency
associated to the system of identical particles Eq. (\ref{18}) is exactly
the same as for distinguishable particles Eq. (\ref{19}), GRWP have chosen $%
\zeta \sim 10^{-30}~cm^3s^{-1}$. %
%%%%%%%%%%%%%%%%%%%%%%%%%%%%%%%%%%%%%%%%%%%%%%%%%%%%%%%%%%%%%%%%%%%%%%%%%%%%%%%%%%%%%%%

\section{Decoherence from the Cosmic Background Radiation}

%%%%%%%%%%%%%%%%%%%%%%%%%%%%%%%%%%%%%%%%%%%%%%%%%%%%%%%%%%%%%%%%%%%%%%%%%%%%%%%%%%%%%%
%
Our approach uses the stochastic dynamical equation (\ref{4}), where we
identify the continuous component (in frequency space) of the operator
responsible for the interaction of the $N$-particle system to the CBR as

\begin{equation}
{\bf Z}(\Omega )\equiv \sum_{k=1}^N\left( A(\Omega ){\bf a}_k^{\dagger
}+A^{\dagger }(\Omega ){\bf a}_k\right) ~,\qquad {\bf a}%
_k=(a_{k,x},a_{k,y},a_{k,z})~.  \label{25}
\end{equation}
where
\begin{equation}
{\bf a}_k=\frac 1{\sqrt{2\hbar m\omega }}\left( m\omega {\bf q}_k+i{\bf p}%
_k\right) ,  \label{252}
\end{equation}
and ${\bf a}_k^{\dagger }$ is its hermitian conjugate $\left( \left[
a_{k,i},a_{k^{\prime },j}^{\dagger }\right] =\delta _{k,k^{\prime }}\delta
_{i,j},i=x,y,z\right) $, ${\bf q}_k$ and ${\bf p}_k$ are respectively
position and momentum operators of the $k^{th}$ particle of mass $m$. $\hbar
\omega $ is a characteristic energy of the system of particles associated to
the quantum fluctuation of the CM. The operators $A^{\dagger }(\Omega
),A(\Omega )$ stand for the creation and annihilation of a quantum of energy
$\hbar \Omega $ from the environment. The coupling parameter is defined by
the continuous stochastic Wiener process ${\bf B}(\Omega )$ satisfying
\begin{mathletters}
\label{26}
\begin{eqnarray}
\overline{d{\bf B}(\Omega )} &=&0,  \label{26a} \\
\overline{dB_i(\Omega )dB_j(\Omega ^{^{\prime }})} &=&\gamma (\Omega )\delta
_{i,j}\delta (\Omega -\Omega ^{^{\prime }})dt,  \label{26b}
\end{eqnarray}
with $\gamma (\Omega )=\Lambda \Gamma (\Omega )$ accounting for a strength
parameter $\Lambda $ and a frequency distribution function $\Gamma (\Omega )$%
. Note that $\Gamma (\Omega )$ refers to the effective frequency
distribution of the CBR photons which interact with the system of particles
at energy around $\hbar \omega $. We next consider that the system of
particles and CBR interacting almost resonantly with Lorentzian spectrum
\end{mathletters}
\begin{equation}
\Gamma (\Omega )=\frac 1\pi \frac{\tau _c}{\tau _c^2(\Omega -\omega )^2+1}.
\label{27}
\end{equation}
In view of Eq. (\ref{27}) it follows from the Fourier transform of Eq. (\ref
{26b}) that
\begin{equation}
\overline{dB_i(t)dB_j(t^{\prime })}=\frac \Lambda {2\pi }\mathop{\rm e}%
\nolimits^{i\omega (t-t^{^{\prime }})}\mathop{\rm e}\nolimits^{-(t-t^{^{%
\prime }})/\tau _c}dt,  \label{28}
\end{equation}
where the correlation time $\tau _c$ defines the memory time over which the
stochastic function changes appreciably. From Eq. (\ref{28}) we conclude
that when considering $\tau _c$ extremely short, {\it i.e.}, much less than
all other times of interest (evolution of the particle system) so that in a
good approximation $\overline{dB_i(t)dB_j(t^{\prime })}\thicksim \delta
(t-t^{^{\prime }})dt$, the system is Markovian. Through Eqs. (\ref{26a}) and
(\ref{26b}) the physical stochastic differential equation (\ref{4}) reads
\begin{eqnarray}
d|\Psi_{N+CBR}\rangle &=&\left\{ -\frac i\hbar H_{N+CBR}dt+\int d\Omega
\sum_{k=1}^{N}\left( A(\Omega ){\bf a}_k^{\dagger }+A^{\dagger }(\Omega )%
{\bf a}_k\right) \cdot d{\bf B}(\Omega )\right.  \nonumber \\
&&-\left. \frac \Lambda 2\int d\Omega \Gamma (\Omega )\left[
\sum_{k=1}^{N}\left( A(\Omega ){\bf a}_k^{\dagger }+A^{\dagger }(\Omega )%
{\bf a}_k\right) \right] ^2dt\right\} |\Psi _{N+CBR}\rangle .  \label{29}
\end{eqnarray}
It must be emphasized that Eq. (\ref{29}) describes the evolution of the
state vector of system of $N$ particles and CBR differently from the
stochastic differential equation in the CSL model. The Hamiltonian $H_{N+CBR}
$ in this equation describes the free evolution of the system of particles
and CBR, while the two remaining terms account for the stochastic
interaction between the CBR and partices.

By defining both, the Wiener process $d{\bf B}$ and the operator ${\bf Z}$
depending on the CBR frequency space, the positional space will not be
anymore privileged with respect to the momentum space, as occurs in the CSL
model. We now proceed to the separation of the CM motion of the modified
dynamical equation (\ref{29}). The substitution of the operators ${\bf a}%
_k^{\dagger },{\bf a}_k$ as position and momentum operators ${\bf p}_k,{\bf q%
}_k$, permits us to express Eq. (\ref{29}) in terms of the CM coordinates $%
{\bf Q=}\frac 1N\sum_k{\bf q}_k$ and ${\bf P}=\sum_k{\bf p}_k$ as
\begin{eqnarray}
d|\Psi _{N+CBR}\rangle &=&\left\{ -\frac i\hbar H_{N+CBR}dt+\int d\Omega
\left( A(\Omega) {\bf X}^{\dagger }+A^{\dagger }(\Omega ){\bf X}\right)
\cdot d{\bf B}(\Omega )\right.  \nonumber \\
&-&\left. \frac \Lambda 2\int d\Omega \Gamma (\Omega )\left( A(\Omega ){\bf X%
}^{\dagger }+A^{\dagger }(\Omega ){\bf X}\right) ^2dt\right\} |\Psi_{N+CBR}
\rangle .  \label{31}
\end{eqnarray}

where the operator ${\bf X}$ accounting for the macroscopic object reads
\begin{equation}
{\bf X}=\frac 1{\sqrt{2\hbar m\omega }}\left( Nm\omega {\bf Q}+i{\bf P}%
\right) ,  \label{32}
\end{equation}
while ${\bf X}^{\dagger }$ is its hermitian conjugate. These operators
satisfy the commutation relation $[X_i,X_j^{\dagger }]=N\delta _{i,j}\hat{1}$%
. As mentioned earlier the coupling constant of the interaction between the
CBR and the system of particles defines a characteristic time $\Lambda ^{-1}$
which is associated to the net effect of the random pseudo-``potential''
described by the last two terms on the right-hand side of Eq. (\ref{31}).

As the stochastic operator in Eq. (\ref{31}) automatically acts only on the
joint wave vector of the CM degree of freedom and the CBR $|\Psi
_{CM+CBR}\rangle $, the separately normalized state vectors $|\Psi
_{CM+CBR}\rangle $ and $\left| \phi _{\{{\bf r}_i\}}\right\rangle $, the
latter for the internal degrees of freedom, satisfy the equations
\begin{mathletters}
\label{320}
\begin{eqnarray}
d|\Psi _{CM+CBR}\rangle &=&\left[ -\frac i\hbar H_{CM+CBR}dt+\int d\Omega
\left( A(\Omega ){\bf X}^{\dagger }+A^{\dagger }(\Omega ){\bf X}\right)
\cdot d{\bf B}(\Omega )\right.  \nonumber \\
&&\left. -\frac \Lambda 2\int d\Omega \Gamma (\Omega )\left( A(\Omega ){\bf X%
}^{\dagger }+A^{\dagger }(\Omega ){\bf X}\right) ^2dt\right] |\Psi
_{CM+CBR}\rangle ,  \label{320a} \\
d|\phi _{\{{\bf r}_i\}}\rangle &=&-iH_{_{\{{\bf r}_i\}}}|\phi _{\{{\bf r}%
_i\}}\rangle dt.  \label{320b}
\end{eqnarray}
It should be noted that the above Eqs. (\ref{320a}) and (\ref{320b}),
differently from those in the CSL model (Eqs. (\ref{17a}) and (\ref{17b})),
involve also the CBR degrees of freedom. As will be shown later, the present
approach in the low temperature limit allows obtaining separately the
normalized wave functions for the system of particles, $|\Psi _{CM}\rangle $
and $\left| \phi _{\{{\bf r} _i\}}\right\rangle $, satisfying equations
similar to those in the CSL. Next, from Eqs. (\ref{6}) and (\ref{31}) the
statistical operator $\rho _{N+CBR}=\overline{|\Psi _{N+CBR}\rangle \langle
\Psi _{N+CBR}|}$ reads
\end{mathletters}
\begin{eqnarray}
\frac{d\rho _{N+CBR}}{dt} &=&-\frac i\hbar [H_{N+CBR},\rho _{N+CBR}]-\frac %
\Lambda 2\int d\Omega \Gamma (\Omega )\left[ \left\{ \left( A(\Omega ){\bf X}%
^{\dagger }+A^{\dagger }(\Omega ){\bf X}\right) ^2,\rho _{N+CBR}\right\}
\right.  \nonumber \\
&&-\left. 2\left( A(\Omega ){\bf X}^{\dagger }+A^{\dagger }(\Omega ){\bf X}%
\right) \cdot \rho _{N+CBR}\left( A(\Omega ){\bf X}^{\dagger }+A^{\dagger
}(\Omega ){\bf X}\right) \right] ,  \label{34}
\end{eqnarray}
which is a pre-master equation in that it contains operators from both, the $%
N$ particles system and the CBR, allowing to calculate correlations between
operators of system of particles and CBR. However, since we only have at our
disposal the statistical properties of the CBR field, the obvious procedure
is to trace over the CBR degrees of freedom, considered thermalized at
temperature $T$, which leads to the reduced density operator of the system
of particles only, containing the average number of photons of the CBR as a
parameter.

Back to Eqs. (\ref{28}), when the following assumptions are met: i) a short
correlation time $\tau _c$ ($\ll \Lambda ^{-1}$), leading to the Markovian
approximation; ii) the interaction between the system of particles and CBR
is sufficiently small (exactly the purpose at hand), the density operator of
the global system can be written as $\rho _{N+CBR}(t)=\rho _{N}(t)\otimes
\rho _{CBR}(t)+\rho _{correl}(t)$, where the correlation term $\rho _{correl}
$ can be neglected \cite{cohen}. By considering the thermalized CBR density
operator $\rho _{CBR}=\exp \left( -\beta H_{CBR}(A^{\dagger },A)\right) /%
{\rm Tr}\left[ \exp \left( -\beta H_{CBR}(A^{\dagger },A)\right) \right] $,
with $\beta =k_BT$, $k_B$ being the Boltzmann's constant and $T$ the CBR
temperature, we find the master equation for the N-particle system
\begin{eqnarray}
\frac{d\rho _{N}}{dt} &=&-\frac i\hbar [H_N,\rho _{N}]-\frac \Lambda 2\int
d(\Omega )\Gamma (\Omega )\left\{ [{\bf X}^{\dagger }{\bf .,X}\rho
_{N}]+[\rho _{N}{\bf X}^{\dagger }{\bf .,X}]\right.  \nonumber \\
&&\left. +~\langle n\rangle _\Omega \left( [{\bf X}^{\dagger }\cdot ,[{\bf X}%
,\rho _{N}]]+[{\bf X}\cdot ,[{\bf X}^{\dagger },\rho _{N}]]\right) \right\} ,
\label{35}
\end{eqnarray}
where $\rho _{N}$ is the {\em reduced density operator} of the N-particle
system and $\langle n\rangle _\Omega =1/\left( \exp (\beta \hbar \Omega
)-1\right)$ is the thermal averaged photon number.

As time goes on, it is expected that the stochastic coupling induces the $N$%
-particle system to a thermal equilibrium with the CBR. By evaluating the
rate of energy change between the system and the CBR we shall estimate the
strength parameter $\Lambda $ and improve our understanding about the nature
of this stochastic coupling. In order to estimate the energy mean-value let
us consider the mean value of a generic dynamical variable ${\cal V}$ whose
equation of motion is obtained through Eq.(\ref{35}) as
\begin{eqnarray}
\frac{d\langle {\cal V}\rangle }{dt} &=&-\frac i\hbar {\rm tr}\left( [{\cal V%
},H_N]\rho _{N}\right) -\frac \Lambda 2\int d(\Omega )\Gamma (\Omega ){\rm Tr%
}\left\{ \left[ [{\cal V},{\bf X}^{\dagger }]\cdot {\bf X}+{\bf X}^{\dagger
}\cdot [{\bf X,}{\cal V}]\right. \right.  \nonumber \\
&&\left. \left. +\langle n\rangle _\Omega \left( [[{\cal V},{\bf X}^{\dagger
}]\cdot ,{\bf X}]+[[{\cal V},{\bf X}]\cdot ,{\bf X}^{\dagger }]\right)
\right] \rho _{N}\right\} ,  \label{36}
\end{eqnarray}

By applying Eq.(\ref{36}) to the position and momentum variables
consecutively, we observe that not only the pure Schr\"{o}dinger evolution
is modified but also the results from the CSL model, such that the equations
of motion become
\begin{mathletters}
\label{37}
\begin{eqnarray}
\frac{d\langle {\bf Q}\rangle }{dt} &=&\frac 1M\langle {\bf P}\rangle -\frac %
12N\Lambda \langle {\bf Q}\rangle ,  \label{37a} \\
\frac{d\langle {\bf P}\rangle }{dt} &=&-\frac 12N\Lambda \langle {\bf P}%
\rangle .  \label{37b}
\end{eqnarray}
These equations lead to the results $\langle {\bf P}\rangle _t=\exp \left( -%
\frac 12N\Lambda t\right) \langle {\bf P}\rangle _s$ and $\langle {\bf Q}%
\rangle _t=\exp \left( -\frac 12N\Lambda t\right) \langle {\bf Q}\rangle _s$%
, where the subscript $s$ indicates the pure Schr\"{o}dinger evolution: $%
\langle {\bf P}\rangle _s=\langle {\bf P}\rangle _{t=0}$ and $\langle {\bf Q}%
\rangle _s=\langle {\bf Q}\rangle _0+\langle {\bf P}\rangle _{t=0}/M\;t$.
For ${\cal V}={\bf Q}^2,{\bf Q}\cdot{\bf P}+{\bf P}\cdot{\bf Q}$ and ${\bf P}%
^2$ successively, the equations of motion for the mean values become,
respectively,
\end{mathletters}
\begin{mathletters}
\label{38}
\begin{eqnarray}
\frac{d\langle {\bf Q}^2\rangle }{dt} &=&\frac 1M\langle {\bf Q \cdot P}+%
{\bf P \cdot Q}\rangle -N\Lambda \langle {\bf Q}^2\rangle +\frac{3\hbar
\Lambda }{2m\omega }\int d\Omega \Gamma (\Omega )\left( 1+2\langle n\rangle
_\Omega \right) ,  \label{38a} \\
\frac{d\langle {\bf Q \cdot P}+{\bf P \cdot Q}\rangle }{dt} &=&\frac 2M%
\langle {\bf P}^2\rangle -N\Lambda \langle {\bf Q \cdot P}+{\bf P \cdot Q}%
\rangle ,  \label{38b} \\
\frac{d\langle {\bf P}^2\rangle }{dt} &=&-N\Lambda \langle {\bf P}^2\rangle +%
\frac{3N^2\Lambda m\hbar \omega }2\int d\Omega \Gamma (\Omega )\left(
1+2\langle n\rangle _\Omega \right) ~,  \label{38c}
\end{eqnarray}
which differ from the pure Schr\"{o}dinger evolution since $\Lambda \neq 0$.
%
%%%%%%%%%%%%%%%%%%%%%%%%%%%%%%%%%%%%%%%%%%%%%%%%%%%%%%%%%%%%%%%%%%

\section{The Master Equation and It\^{o} Dynamics}

%%%%%%%%%%%%%%%%%%%%%%%%%%%%%%%%%%%%%%%%%%%%%%%%%%%%%%%%%%%%%%%%%%%%%%%%%%%
%
It will be useful to remind the conventional treatment of the problem of
interaction of a N-particle system with the reservoir (R). Under the
Hamiltonian $H=H_N+H_{R}+V$, $V$ being the interaction between both systems,
the reduced density operator of the N-particle system, $\rho _{N}(t)={\rm Tr}%
_R\left[ \rho _N(t)\right] $, evolves, up to the second order in the
interaction, according to the generalized master equation \cite{blum}
\end{mathletters}
\begin{equation}
\frac{d\rho _{N}(t)}{dt}=-\frac i\hbar [H_N,\rho _{N}(t)]-\frac 1{\hbar ^2}%
{\rm Tr}_R\int_0^t[V, \mathop{\rm e}\nolimits^{-iL_0(t-t\acute{})}[V,\rho
_{N}(t\acute{})\rho _R]]dt\acute{},  \label{381}
\end{equation}
where $L_0(\cdot )\equiv [H_N+H_R,\cdot ]$ is the Liouvillian operator of
the free Hamiltonian. The second term in Eq.(\ref{381}), acting as a source
of noise for the system and also as a sink (or source) of energy, is
responsible for the irreversibility of the process and the loss of coherence
in $\rho _{N}(t)$. As such, the It\^{o} calculus is justified when the
stochastic terms are introduced into the Schr\"{o}dinger equation. So, the
CBR is responsible for the variation of the mean energy of the system and
the increase of entropy. As shown by Isar {\it et al. }\cite{isar}, choosing
conveniently the interaction term $V$ it is possible to obtain Eq. (\ref{35}%
) (the Lindblad form) from Eq. (\ref{381}).

It is worth noting that the master equation (\ref{35}) can be written as
\begin{equation}
\frac{d\rho _{N}(t)}{dt}=-\frac i\hbar [H_N,\rho _{S}(t)]+\sum_{n=1}^2{\cal S%
}[c_n]\rho _{N}(t),  \label{382}
\end{equation}
where the superoperator ${\cal S}[c_n]$ is defined as
\begin{equation}
{\cal S}[c_n]\rho _{N}=c_n.\rho _{N}c_n^{\dagger }-\frac 12\left\{
c_n^{\dagger }.c_n,\rho _{N}\right\} ,  \label{383}
\end{equation}
with $c_1=\left[ \Lambda \int d\Omega \Gamma (\Omega )\langle n\rangle
_\Omega \right] ^{1/2}{\bf X}^{\dagger }$ and $c_2=\left[ \Lambda \int
d\Omega \Gamma (\Omega )\left( 1+\langle n\rangle _\Omega \right) \right]
^{1/2}{\bf X}$. Written as in Eq. (\ref{382}) our master equation resembles
the Lindblad form for the decay of a mode of the eletromagnetic field inside
a cavity \cite{scully}.

In summary, we have assumed {\em ad-hoc} that the evolution of the system of
particles in its way from quantum to classical dynamics, under the influence
of the CBR, is described by an It\^{o} stochastic equation. However, here we
showed that the usual master equation formalism can be viewed as a
subdynamics of the It\^{o} dynamics, without any need to use perturbation
methods as is done in the conventional derivation. %
%%%%%%%%%%%%%%%%%%%%%%%%%%%%%%%%%%%%%%%%%%%%%%%%%%%%%%%%%%%%%%%%%%%%%%%%%%%%%%%%%%%%%%

\section{The strength parameter}

%%%%%%%%%%%%%%%%%%%%%%%%%%%%%%%%%%%%%%%%%%%%%%%%%%%%%%%%%%%%%%%%%%%%%%%%%%%%%%%%%%%%%%
%
Back to the equations of motion (\ref{38}), their solutions are
\begin{mathletters}
\label{39}
\begin{eqnarray}
\langle {\bf Q}^2\rangle &=&\langle {\bf Q}^2\rangle _s\mathop{\rm e}%
\nolimits^{-N\Lambda t}-\frac{3{\cal I}\hbar \omega }M\left[ \frac t{%
N\Lambda }\left( 1-\frac{N\Lambda t}2\right) \mathop{\rm e}%
\nolimits^{-N\Lambda t}-\left( \frac 1{N^2\Lambda ^2}+\frac 1{2\omega ^2}%
\right) \left( 1-\mathop{\rm e}\nolimits^{-N\Lambda t}\right) \right] ,
\label{39a} \\
\langle \left\{ {\bf Q},{\bf P}\right\} \rangle &=&\langle \left\{ {\bf Q},%
{\bf P}\right\} \rangle _s\mathop{\rm e}\nolimits^{-N\Lambda t}-3{\cal I}%
\hbar \omega \left[ t\mathop{\rm e}\nolimits^{-N\Lambda t}-\frac 1{N\Lambda }%
\left( 1-\mathop{\rm e}\nolimits^{-N\Lambda t}\right) \right] ,  \label{39b}
\\
\langle {\bf P}^2\rangle &=&\langle {\bf P}^2\rangle _s\mathop{\rm e}%
\nolimits^{-N\Lambda t}+\frac{3{\cal I}Nm\hbar \omega }2\left( 1-\mathop{\rm
e}\nolimits^{-N\Lambda t}\right) \;,  \label{39c}
\end{eqnarray}
where
\end{mathletters}
\begin{mathletters}
\label{391}
\begin{eqnarray}
\langle {\bf Q}^2\rangle _s &=&\langle {\bf Q}^2\rangle _0+\frac 1M\left(
\langle \left\{ {\bf Q,P}\right\} \rangle _0t+\frac 1M\langle {\bf P}%
^2\rangle _0t^2\right) ,  \label{391a} \\
\langle \left\{ {\bf Q,P}\right\} \rangle _s &=&\langle \left\{ {\bf Q,P}%
\right\} \rangle _0+\frac 2M\langle {\bf P}^2\rangle _0t,  \label{391b} \\
\langle {\bf P}^2\rangle _s &=&\langle {\bf P}^2\rangle _0.  \label{391c}
\end{eqnarray}
The effect of the CBR temperature is present in the integral ${\cal I}=\int
d\Omega \Gamma (\Omega )\left( 1+2\langle n\rangle _\Omega \right) $. It is
worth noting that the time evolution of the operators in Eqs. (\ref{39})
does not show the additive property with respect to the Schr\"{o}dinger
terms as obtained in the CSL model. As a consequence, Eq. (\ref{39c})
differs from the corresponding one in the CSL model, Eq. (\ref{20b}),
because instead of the diffusion, inducing a steady increase of the mean
value of the kinetic energy, the present model exhibits, asymptotically,
thermalization due to the CBR,
\end{mathletters}
\begin{equation}
\langle K\rangle =\left( \langle K\rangle _s-K_{eq}\right)
%TCIMACRO{\limfunc{e} }
%BeginExpansion
\mathop{\rm e}%
%EndExpansion
\nolimits^{-N\Lambda t}+K_{eq},  \label{40}
\end{equation}
where the equilibrium kinetic energy reads $K_{eq}=3{\cal I}\hbar \omega /4.$
So, $\omega $ is a characteristic frequency proportional to the thermalized
mean kinetic energy of the CM.

As mentioned above, in the CSL model the localization of a single particle
of the system is sufficient to localize the whole system; as a consequence,
the CM energy increases linearly with the ``interaction'' parameter $%
N\Lambda t$. However, from Eq. (\ref{40}) we conclude that the stochastic
coupling accounts for a CM energy which grows or decays exponentially with $%
N\Lambda t$, depending on the negative or positive value for $\langle
K\rangle _s-K_{eq}$, respectively.

In order to estimate the strength parameter $\Lambda $, from Eq. (\ref{40})
we assume that the relaxation time follows from relation $(\langle K\rangle
_s-K_{eq})\mathop{\rm e} ^{-N\Lambda \tau _R}\sim K_{eq}$, so that
\begin{equation}
\Lambda \thickapprox \frac 1{N\tau _R}\ln \left( \frac{\langle K\rangle
_s-K_{eq}}{K_{eq}}\right) .  \label{41}
\end{equation}
For a system of $N\approx 10^{23}$ particles initially at room temperature
the equipartition energy theorem gives a mean kinetic energy $\langle
K\rangle _s\thicksim 10^9$ergs. The integral ${\cal I}$ accounting for the
effect of the temperature of the CBR has been estimated in the Appendix for $%
\beta \tau _c\ll \hbar $, with $\omega \tau _c\lesssim 1$. The result ${\cal %
I}$ $\thicksim 1+2\langle n\rangle _\omega $, holds for both, low- and
high-frequency regimes. So, we find for the equilibrium energy at
low-frequency regime ($\hbar \omega \ll k_BT$, so that $\langle n\rangle
_\omega \sim k_BT/\hbar \omega $), $K_{eq}\thicksim k_BT\thicksim
10^{-16}ergs$. At high-frequency regime ($\hbar \omega \gg k_BT$), the
equilibrium energy obeys $K_{eq}\gg k_BT$. (We are referring to low- and
high-frequency regimes since the nowadays CBR temperature, $T\approx 3$K is
assumed). Taking $K_{eq}$ at low-frequency regime (in fact, due to the $\ln$
function, choosing $K_{eq}$ in low- or high- frequency will not change
appreciably the value of $\Lambda $), and the relaxation time $\tau _R$ of
the order of the age of the Universe, about $10^{16}s$ (what seems to be
reasonable when considering, as obtained below, such a small coupling of the
system to the CBR), we get
\begin{equation}
\Lambda \thickapprox 10^{-38}s^{-1},  \label{42}
\end{equation}
a value to be compared with the above-mentioned coupling in the CSL model $%
\zeta \sim 10^{-30}~cm^3s^{-1}$. Thus, the parameter $\Lambda $ is of the
order of the upper limit of the excitation rate for nucleons estimated by
Pearle and Squires \cite{philip}, by comparison with neutrino-induced
process. As already pointed out, such a value hardly affects the dynamics of
a microscopic particle. %
%%%%%%%%%%%%%%%%%%%%%%%%%%%%%%%%%%%%%%%%%%%%%%%%%%%%%%%%%%%%%%%%%%%%%%%%%%%%%%%%%%%%%%%%%%%%%%%%%%

\section{Wave-packets reduction rates}

%%%%%%%%%%%%%%%%%%%%%%%%%%%%%%%%%%%%%%%%%%%%%%%%%%%%%%%%%%%%%%%%%%%%%%%%%%%%%%%%%%%%%%%%%%%%%
%
Back to Eq. (\ref{35}), in the CM positional representation, the density
matrix $\rho_{N} ({\bf Q},{\bf Q^{\prime }})$ evolves according to the
differential equation
\begin{eqnarray}
\frac{\partial \rho _N({\bf Q},{\bf Q^{\prime },t})}{\partial t} &=&\left\{ -%
\frac \hbar {2iM}\left( \frac{\partial ^2}{\partial {\bf Q}^2}-\frac{%
\partial ^2}{\partial {\bf Q^{\prime }}^2}\right) -{\cal D}\left[ \left(
{\bf Q}-{\bf Q^{\prime }}\right) ^2-\frac{\hbar ^2}{\left( M\omega \right) ^2%
}\left( \frac \partial {\partial {\bf Q}}+\frac \partial {\partial {\bf %
Q^{\prime }}}\right) ^2\right] \right.  \nonumber \\
&-&\left. \frac 12N\Lambda \left[ \left( {\bf Q}\cdot \frac \partial {%
\partial {\bf Q^{\prime }}}+{\bf Q^{\prime }}\cdot \frac \partial {\partial
{\bf Q}}\right) -1\right] \right\} \rho _N({\bf Q},{\bf Q^{\prime },t}).
\label{43}
\end{eqnarray}
The first term on the right-hand side comes from the commutator in Eq. (\ref
{35}), the terms multiplied by the diffusion constant ${\cal D}= NM\Lambda
\omega \left( 1+2\langle n\rangle _\omega \right)/4\hbar $ (as well as the
remaining terms, which are independent of temperature) account for the
fluctuations (or random kicks) and for the energy changes due to the
stochastic coupling, respectively.

To analyze the wave-packet reduction rates we will not consider Eq. (\ref{43}%
) in detail, since the effect of the second term on quantum superposition
will be of much greater interest \cite{zurek}. For a brief estimation of the
off-diagonal matrix elements of Eq. (\ref{43}) will decay exponentially as
\begin{equation}
\langle {\bf Q}|\rho _S(t)|{\bf Q^{\prime }}\rangle =\mathop{\rm e}
\nolimits^{-\zeta t}\langle {\bf Q}|\rho _S(0)|{\bf Q^{\prime }}\rangle ,
\label{46}
\end{equation}
where $\zeta ={\cal D}(\Delta {\bf Q})^2$ and $(\Delta {\bf Q})^2=\left(
{\bf Q}-{\bf Q^{\prime }}\right) ^2$. It follows from Eq. (\ref{46}) that
the quantum coherence of a macroscopic system will disappear on a
decoherence time scale
\begin{equation}
\tau _D\approx \frac 1{{\cal D}(\Delta {\bf Q})^2}=\frac 1{\left( 1+2\langle
n\rangle _\omega \right) }\frac \hbar {NM\Lambda \omega (\Delta {\bf Q})^2}.
\label{47}
\end{equation}

Analyzing Eq. (\ref{47}) in terms of the CBR temperature, it is interesting
to note that in the low-temperature limit (nowadays universe, $T\sim 3K$),
{\it i.e.}, $\langle n\rangle _\omega \rightarrow 0$, the number of
particles $N$ plays a crucial role in the decoherence process induced by the
CBR. In the high-temperature limit, i.e, $\langle n\rangle _\omega
\rightarrow \infty $ (the early universe in the present model), we conclude
that Eq. (\ref{47}) leads from quantum to classical physics even a system
composed by a small number of particles. This is a key result which help
supporting the assumptions considered in the present model.

Let us now estimate the decoherence time for both, a macroscopic and a
microscopic object in nowadays Universe, i.e, $T\sim 3K$. In order to
compare our results with that presented in literature, we consider the
low-frequency regime, such that Eq. (\ref{47}) reduces to
\begin{equation}
\tau _D\approx \frac 1{{\cal D}(\Delta {\bf Q})^2}= \frac {\hbar^2}{2
N\Lambda M k_BT(\Delta {\bf Q})^2}.  \label{474}
\end{equation}
By considering a system of $N$ ($\sim 10^{23})$ hydrogen atoms with mass $%
M\thickapprox 1$g and separation $\Delta {\bf Q}\thickapprox 1$cm, quantum
coherence would be destroyed in $\tau _D\approx 10^{-24}s$. Such a value
turns to be significantly smaller than the one obtained by GRWP, $\lambda
_{CM}\thickapprox 10^{-7}s$ , Eq. (\ref{19}), and comparable with that
obtained through the linear response model of the Caldeira and Leggett (CL)
\cite{amir}, where, also at low-frequency regime, $\tau _D/\tau
_R\thickapprox \hbar ^2/2mk_BT(\Delta {\bf Q})^2$, $\tau _R$ being a
relaxation time. For the above-mentioned system of $N$ atoms, and assuming $%
\tau _R\thickapprox 10^{16}s$, as we have done to obtain $\Lambda $, Eq. (%
\ref{41}), we get from CL model $\tau _D\thickapprox 10^{-23}s$. So, Eq. (%
\ref{47}), and consequently Eq. (\ref{474}), arise from a theory that,
despite assuring the essential character of the GRWP model, gives a more
realistic value for the decoherence time of a macroscopic object.

As far as a microscopic object is concerned, for example a single atom, $%
m\thickapprox 10^{-24}g$ on atomic scale $\Delta {\bf Q}\approx 10^{-8}cm$,
we observe the persistence of quantum coherence since $\tau _D\approx
10^{41}s$. Finally, we note that when considering a tiny Weber bar \cite
{braginsky,zurek}, $\Delta {\bf Q}\approx 10^{-19}m$, at cryogenic
temperatures, $T\approx 10^{-3}K$, we also observe the persistence of
quantum coherence from Eq. (\ref{47}), as should be expected.

Back to Eq. (\ref{46}), when interpreting the exponential damping factor $%
\zeta $ by the light of the CSL model (Eqs. (\ref{18}) and (\ref{19})), we
conclude that the strength $\Lambda $ plays the role of a microscopic
frequency hitting parameter. %
%%%%%%%%%%%%%%%%%%%%%%%%%%%%%%%%%%%%%%%%%%%%%%%%%%%%%%%%%%%%%%%%%%%%%%%%%%%%%%%%%

\section{The CM and Internal Motion}

%%%%%%%%%%%%%%%%%%%%%%%%%%%%%%%%%%%%%%%%%%%%%%%%%%%%%%%%%%%%%%%%%%%%%%%%
%
By construction we assumed that the CBR acts only on the CM coordinates of
the system of particles. Such assumption automatically decouples the
dynamics of the collective and internal motions in the master equation (\ref
{35}). Next, we show that even the vector state dynamics for the CM and the
internal motion decouple, as in the CSL model. Of course, our analysis will
be restricted to the low temperature limit where, as obtained in Eq. (\ref
{47}), the macroscopic character of the system becomes really important due
to the number of particles $N$. In this limit Eq. (\ref{35}) simplifies to
\begin{equation}
\frac{d\rho _N}{dt}=-\frac i\hbar [H_N,\rho _N]+\Lambda {\bf X}\cdot\rho _N%
{\bf X}^{\dagger }-\frac \Lambda 2\left\{ {\bf X^{\dagger }\cdot X},\rho
_N\right\} .  \label{470}
\end{equation}
The stochastic differential equation for the state vector of the system of
particles which leads to Eq. (\ref{470}) can be written as
\begin{equation}
d|\Psi _N\rangle =\left( -\frac i\hbar H_Ndt+{\bf X}\cdot d{\bf W}-\frac %
\Lambda 2{\bf W{\dagger }\cdot W}^dt\right) |\Psi _S\rangle ,  \label{471}
\end{equation}
now with the Wiener process $\overline{dW_i}=0,~~\overline{dW_idW_j}=\Lambda
\delta _{ij}dt.$

The assumption made in the CSL model, that the set $\{{\bf r}_i\}$ in Eq. (%
\ref{13}) does not contain macroscopic variables, implies that the state
vector for the macroscopic object factorizes as $\Psi _N(\{{\bf q}_k\})=\psi
_{CM}({\bf Q})\phi _{int}(\{{\bf r}_i\})$. The additional assumption that
the CM motion is decoupled from the internal degrees of freedom means that
the Hamiltonian must be written as a sum of two terms, $H_N=H_{CM}+H_{int}$
\cite{gpr}. Under these assumptions the It\^{o} calculus, $d\Psi _N=d(\psi
_{CM}\phi _{int})=(d\psi _{CM})\phi _{int}+ \psi _{CM}(d\phi _{int})+
\overline{(d\psi _{CM}(d\phi _{int})}$, shows that the wave functions $\psi
_{CM}({\bf Q})$ and $\phi _{int}({{\bf r}_i})$, similarly to Eqs. (\ref{17a}%
) and (\ref{17b}), satisfy equations
\begin{mathletters}
\label{472}
\begin{equation}
d|\psi _CM\rangle =\left( -\frac i\hbar H_{CM}dt+{\bf X}\cdot d{\bf W}-\frac %
\Lambda 2{\bf W^{\dagger }\cdot W}dt\right) |\Psi _{CM}\rangle ,
\label{472a}
\end{equation}
\begin{equation}
d|\phi _{int}\rangle =-\frac i\hbar H_{int} |\phi _{int}\rangle dt.
\label{472b}
\end{equation}
The stochastic terms do not affect the internal structure of the system of
particles, {\it i.e.}, nothing changes in the Schr\"{o}dinger dynamics of
microscopic particles. It is worth noting that in the CSL model the
additional assumption of a large enough localization width parameter
(besides of an internal wave function independent of macroscopic variables)
is necessary to decouple the dynamics of $\psi _{CM}$ from $\phi _{int}$. In
fact, as shown in Ref. \cite{miled}, a width parameter of order of atomic
size leads to the breakdown of the translational symmetry of the system and
the interaction between the CM and the relative coordinates ({\it i.e.}, $%
H=H_{CM}+H_{int}+V $), has to be taken into account. However, in the present
model, since we have assumed that the CBR acts only on the CM coordinates of
the system of particles, no additional conjectures was requested about the
random operator ${\bf Z}(\Omega )$, Eq. (\ref{25}), to achieve the
remarkable result of the CM decoupling from the internal motion, as if the
stochastic terms in Eq. (\ref{35}) were absent. The operator ${\bf Z}(\Omega
)$ has thus the advantage of not needing additional conjectures about the
width parameter of the localization process. %
%%%%%%%%%%%%%%%%%%%%%%%%%%%%%%%%%%%%%%%%%%%%%%%%%%%%%%%%%%%%%%%%%%%%%%%%%%%%%%%%%%%%%%

\section{Decoherence and Entropy}

%%%%%%%%%%%%%%%%%%%%%%%%%%%%%%%%%%%%%%%%%%%%%%%%%%%%%%%%%%%%%%%%%%%%%%%%%%%%%%%%%%%
%
The decoherence process resulting from the interaction of the state vector
for a macroscopic object with the CBR can be quantified by the rate of
increase of either the linear or the statistical entropy. In terms of the
density matrix, the statistical entropy, a measure of our ignorance, is
defined as \cite{neumann} ${\cal S}_s=-{\rm Tr}\left( \rho \ln \left( \rho
\right) \right) $ (the subscript $s$ refers to {\it statistical}). This
definition does not require that the system be in a thermal equilibrium
state. Alternatively, a good measure of the loss of purity for states of an
evolving open system is based on the increase of the linear entropy
(subscript $l$) \cite{zurek1}
\end{mathletters}
\begin{equation}
{\cal S}_l={\rm Tr}\left( \rho -\rho ^2\right) .  \label{48}
\end{equation}
Next, we estimate the rate of increase of the linear entropy through the
evolution of the density matrix given in the operator form by Eq. (\ref{43}%
). Considering a weak strength parameter $\left( \Lambda \approx 0\right) $
and the state vector remaining approximately pure $\left( {\rm Tr} \rho
^2\approx 1\right) $, up to first order in $\Lambda$ we obtain
\begin{equation}
\dot{{\cal S}}_l=4{\cal D}\left( \langle \left( \Delta {\bf Q}\right)
^2\rangle +\frac 1{\left( Nm\omega \right) ^2}\langle \left( \Delta {\bf P}%
\right) ^2\rangle \right) ,  \label{49}
\end{equation}
where $\langle \left( \Delta {\bf Q}\right) ^2\rangle $ and $\langle \left(
\Delta {\bf P}\right) ^2\rangle ,$ obtained from Eqs. (\ref{39a}) - (\ref
{391c}), stand for the variances of the position and momentum operators and
can be rewritten as function of their initial values $\langle {\bf Q}\rangle
_0$ and $\langle {\bf P}\rangle _0$.

In order to better understand the rate of increase of the linear entropy in
Eq. (\ref{49}), it is worth to compare it with that obtained by Zurek \cite
{zurek1} who used the linear response model of Caldeira and Leggett \cite
{caldeira} (in the high temperature limit). With the above approximations
Zurek obtained $\dot{{\cal S}}_l=4{\cal D}\langle \left( \Delta {\bf Q}%
\right) ^2\rangle $ (for a single oscillator), so that the rate of increase
of linear entropy (in quantum Brownian motion) is proportional to the
dispersion in position coordinate only - the preferred observable singled
out by the interaction Hamiltonian. In our approach, from Eq. (\ref{49}) we
observe that no preferred observable emerge from the dynamic equation (\ref
{35}) (the dispersion in momentum is also present), contrarily even to the
CSL model where the position representation is taken from the outset as
privileged. However, for a large number of particles ($N\gg 1$), Eq. (\ref
{49}) indicates that the dispersion in momentum is considerably smaller when
compared with that in position which, in this situation, emerges as the
preferred observable.

In the weak-coupling limit we integrate Eq. (\ref{49}) replacing the general
evolution in Eq. (\ref{35}) by the free von Neumann equation to obtain
\begin{eqnarray}
{\cal S}_l &=&4{\cal D}\left[ \left( \langle \left( \Delta {\bf Q}\right)
^2\rangle _0+\frac 1{\left( Nm\omega \right) ^2}\langle \left( \Delta {\bf P}%
\right) ^2\rangle _0\right) t+\frac 1{2M}\langle \Delta \left\{ {\bf Q},{\bf %
P}\right\} \rangle _0t^2\right.  \nonumber \\
&+&\left. \frac 1{3M^2}\langle \left( \Delta {\bf P}\right) ^2\rangle
_0t^3\right] ,  \label{50}
\end{eqnarray}
with $\langle \Delta \left\{ {\bf Q,P}\right\} \rangle \equiv \langle
\left\{ {\bf Q,P}\right\} \rangle -2\langle {\bf Q}\rangle \langle {\bf P}%
\rangle $. The dispersions appearing in the Eq.(\ref{50}) are computed for
the pure initial state.

Back to the preferred basis problem, we remind that Zurek considered the
free Heisenberg equations for the oscillator operators ($P,Q$) and obtained
the linear entropy $2{\cal D}\left( \langle \left( \Delta Q\right) ^2\rangle
_0+\frac 1{\left( Nm\omega \right) ^2}\langle \left( \Delta P\right)
^2\rangle _0\right) $ ($N=1$), averaged over one oscillator period. So, this
result corresponds only to the coefficient for the linear time-dependence in
Eq.(\ref{50}), where additional terms as square and cubic time-dependent
behavior also take place. Such a behavior indicates that, in spite of the
large number of particles, for large times the momentum plays an important
role in the problem of the preferred observable because we have considered
the free motion of a $N$-particle system instead of a single harmonic
oscillator. %
%%%%%%%%%%%%%%%%%%%%%%%%%%%%%%%%%%%%%%%%%%%%%%%%%%%%%%%%%%%%%%%%%%%%%%%%%%%%%%%%%

\section{Summary and Conclusions}

%%%%%%%%%%%%%%%%%%%%%%%%%%%%%%%%%%%%%%%%%%%%%%%%%%%%%%%%%%%%%%%%%%%%%%%%%%%%%%%%%%
%
In the GRWP model of continuous dynamical reduction of the state vector it
is assumed that each microscopic constituent of a system of $N$ particles is
subject to a sudden collapse due to a spontaneous random hitting process
consisting in a localization of the wave function of the particle within an
appropriate range \cite{gpr}. In what turns to be a remarkable result the
localization of a single constituent of the system of particles is
sufficient to localize the whole system. Such a spontaneous localization,
considered as a fundamental physical process, induces a steady increase of
the mean energy value of the physical system and so the increase in
temperature per unit time of the universe. When taking into account that the
age of the universe is about $10^{16}s$, the GRWP model leads to a total
temperature increase from the beginning of the universe of $10^{-3}K$, a
value claimed to be comparable with the cosmic background radiation (CBR) of
$3$K.

In the present model for continuous dynamical reduction, also based in a
stochastic differential equation describing a Markovian evolution of state
vectors, the random hitting process in GRWP model is substituted by the
intervention of the CBR. Such a strategy is intended to maintain (i) the
principle of conservation of energy, and (ii) the claim that the Universe
originated from the Big Bang leaving the CBR as a signature. In (i) the
increase or decrease of the CM mean energy of the system of $N$ particles
subject to a stochastic interaction with the CBR, which acts as a reservoir.
In (ii), taking the opposite direction to the GRWP argument (which claim
that the present temperature of the universe comes from the increase of the
total energy arising from the random hitting process), we propose that the
CBR temperature plays an important role in the reduction of the $N$-particle
wave packet. So, we assumed, in agreement with the standard cosmology, that
the Universe has originated from a hot state and cooling during its
expansion, with decreasing mean photon energy. The Planck law for the
thermal average boson number in CBR, indeed the best blackbody known, has
recently been tested by the COBE satellite \cite{cobe}. The temperature of
the CBR, decreasing as the mean photon energy decrease due to the cosmic
expansion makes the mass of the system increasingly more important for the
transition from quantum to classical description. On this basis one can
argue that the quantum nature of the Universe becomes increasingly important
as it is cooling. In fact, for the early Universe, the number of particles
does not play a fundamental role in estimating the decoherence time, where
higher temperatures (by itself) turn the system from micro to macro
dynamics. However, as the Universe becomes cooler the number of particles
becomes increasingly important.

Moreover, the present model leads to realistic results for decoherence
times. While in the GRWP model the value $10^{-7}s$ obtained for a system of
particles to go from micro to macro dynamics seems to be too large, the
value $10^{-24}s$ here obtained for a system of $N$ atoms in the
low-frequency regime is comparable to the decoherence time obtained from the
Caldeira-Leggett model.

As mentioned, whereas the GRWP model requires a {\it privileged positional
space}, in the present model, by construction, the stochastic operator acts
on the CBR spectrum, carrying the same status for both, the position and the
momentum space. The GRWP's result - the wave function collapse of a single
particles induces the collapse of the wave function of the whole system -
was obtained exactly from the choice of the position as a preferred basis.
The same result follows from our model without the choice of the position as
a preferred basis. However, It has to be mentioned that in spite of
attributing the same status for the position and the momentum space, when
analyzing the entropy under the process of decoherence, the position
coordinate still emerges as a preferred basis when considering a system with
a large number of particles $N$. So, the preferred basis is directly related
to the number of particles in the system.

Another interesting feature is that we do not claim for an {\it additional
assumption} to decouple collective from internal motion as the required
large width parameter $\alpha ^{-1/2}\thicksim 10^{-5}cm$ in the GRWP model.
The random operator ${\bf Z}(\Omega )$ here assumed, besides being a more
conventional choice since it is associated to a reservoir (CBR), leads to
the advantage of decoupling the CM and internal motion without additional
assumption beyond that usually assumed for a reservoir.

The random operator describing the interaction between the system and the
CBR carries only one parameter, the strength $\Lambda $, instead of the two
free parameters, as required in the GRWP model ($\alpha ^{-1/2}$ and the
mean frequency $\lambda $). In our model, the coupling of the CBR to the
system, proportional to $\Lambda $, corresponds to the random
pseudo-``potential'' $dh$ \cite{tony} of the GRWP model. As well as the
parameter $\lambda $ in GRWP model, our $\Lambda $ is weak enough in the
sense that it does not affect the dynamics of a unique particle, even in the
case in which its wave function is spatially spread \cite{gpr}.

Finally, we point out that the It\^{o} equation is not derived from a
physical picture of the background and associated scattering processes of
the CBR by the system of particles. Instead of considering a particular
interaction and choose some specific particle property sensible to the
electric and magnetic field of the CBR, we approached the problem by
modeling the interaction by a stochastic coupling, such that the dynamics
could be described by an It\^{o} equation. We have considered an effective
strenght parameter $\Lambda $ accounting for all kind of light-particle
scattering processes. We also stress that our pre-master equation (34) (with
respect to the particles) has still information on both, the system of
particles and the CBR, since it contains operators of both subsystems. This
approach is different from the usual one where for getting a master equation
it is necessary to trace over the environment degrees of freedom, as is done
in the theories of Joos and Zeh and Caldeira-Leggett or even in quantum
optics. In our model it is possible to calculate correlations between
observables of both subsystems. However, we have get rid of CBR degrees of
freedom, Eq. (35), just because the available information on the CBR
subsystem is sparse, consisting of the blackbody radiation distribution
function at 3K. Thus the master equation (35) expressed in the CM positional
representation, Eq. (49), incorporates the similar equations obtained in
both theories, Joos and Zeh and Caldeira-Leggett. The main difference
between the three approaches stem in the nature of the diffusion constant
(DC): In Joos and Zeh the DC originates from the scattering of
electromagnetic waves by small objects; in Caldeira-Leggett it comes from
the fluctuations arising from energy dissipation of the system of interest
to a thermal reservoir. In our model the DC originates from the stochastic
interaction between $N$ particles of mass $m$ and the CBR at temperature $T$%
. %
%%%%%%%%%%%%%%%%%%%%%%%%%%%%%%%%%%%%%%%%%%%%%%%%%%%%%%%%%%%%%%%%%%%%%%%%%%%%%%%
%

\acknowledgments{MCO and NGA thank FAPESP, S\~ao Paulo,
Brazil, for total financial support. MHYM and SSM thank CNPq,
Brazil, for partial financial support. The authors wish to thank
Prof. R. J. Napolitano for helpful discussions.}
 \appendix
%%%%%%%%%%%%%%%%%%%%%%%%%%%%%%%%%%%%%%%%%%%%%%%%%%%%%%%%%%%%%%%%%%%%%%%%%%%5

\section{Calculation of integral ${\cal I}$}

%%%%%%%%%%%%%%%%%%%%%%%%%%%%%%%%%%%%%%%%%%%%%%%%%%%%%%%%%%%%%%%%%%%%%%%%%%%%%%%
%
Due to the normalized Lorentzian spectrum (Eq. (\ref{27})), the integral
accounting for the temperature of the CBR reads ${\cal I}=1+2\int d\Omega
\Gamma (\Omega )\langle n\rangle _\Omega $. Now, since the Planck's
distribution $\langle n\rangle _\Omega $ diverges when $\Omega $ goes to
zero, the same occurs to the remaining integral $\int d\Omega \Gamma (\Omega
)\langle n\rangle _\Omega $. However, as usual, we assume that the spectrum $%
\Gamma (\Omega )$ has its maximum far away from zero in order to cancel the
divergence coming from $\langle n\rangle _\Omega $. In what follows we are
going to estimate under which conditions this approximation is valid.

After the transformations $\Omega \tau _c=x$ and $\gamma =\hbar /k_BT\tau _c$%
, the remaining integral reads
\begin{equation}
\frac 1\pi \int_{-\infty }^{+\infty }dx\frac 1{[x-(\omega \tau
+i)][x-(\omega \tau _c-i)]}\frac 1{%
%TCIMACRO{\func{e} }
%BeginExpansion
\mathop{\rm e}%
%EndExpansion
^{\gamma x}-1},  \label{A1}
\end{equation}
which can be solved in the complex space through Jordan's lema, leading to
the result
\begin{equation}
2i\left\{ \frac 1{2i}\frac 1{%
%TCIMACRO{\func{e} }
%BeginExpansion
\mathop{\rm e}%
%EndExpansion
^{\gamma (\omega \tau _c+i)}-1}+\frac 1\gamma \sum_{n=0}^\infty \left( 1-%
\frac 12\delta _{n,0}\right) \frac 1{[\omega \tau +i(1-\frac{2\pi n}\gamma
)][\omega \tau _c-i(1+\frac{2\pi n}\gamma )]}\right\} .  \label{A2}
\end{equation}
It can be shown that the imaginary term coming from the above result is
zero. Now, denoting $\gamma =p/\xi $, where the parameter $p$ is equal to $%
\hbar \omega /k_BT$ whereas $\xi =\omega \tau _c$, the real term coming from
(\ref{A2}), reads
\begin{equation}
\frac{\cos (\xi /p)%
%TCIMACRO{\func{e} }
%BeginExpansion
\mathop{\rm e}%
%EndExpansion
^\xi -1}{%
%TCIMACRO{\func{e} }
%BeginExpansion
\mathop{\rm e}%
%EndExpansion
^\xi [%
%TCIMACRO{\func{e} }
%BeginExpansion
\mathop{\rm e}%
%EndExpansion
^\xi -2\cos (\xi /p)]+1}-8\pi \frac{p^3}{\xi ^2}\sum_{n=1}^\infty \frac n{%
[1+p^2-(2\pi np/\xi )^2]+(4\pi np^2/\xi )^2}.  \label{A3}
\end{equation}
For large $n$ the second term of (\ref{A3}) reduces to
\begin{equation}
\thicksim \frac{\xi ^2}p\sum_{n=1}^\infty \frac 1{n^3}.  \label{A4}
\end{equation}
The analysis of the above result will be restricted to the condition $\xi
/p\ll 1$, with $\xi \lesssim 1$, under which the sum in (\ref{A4}) can be
disregarded (since even $\xi ^2/p\ll 1)$, and the first term in (\ref{A3})
gives us $1 \mathop{\rm e} ^{\hbar \omega /k_BT}-1)$, in a way that the
Lorentzian distribution $\Gamma (\Omega )$ acts practically as a delta
function ($\delta (\Omega -\omega )$). In fact, the limit$\ \xi \lesssim 1$,
leads to the condition $\omega \lesssim \tau _c^{-1}$, so that the frequency
can be taken far away from zero since, as above-discussed, we are
considering an extremely short correlation time (Markovian approximation).
Under such condition it is expected that the lorentzian function $\Gamma $
acts indeed as a delta function, what means that the action of the reservoir
over the system of particles is restricted to the oscillators whose
frequencies is closely related to $\omega $. So, the problem of how far $%
\omega $ has to be from zero, in order to eliminate the divergence coming
from Planck's distribution when $\omega \rightarrow 0,$ depends exactly on
the lorentzian height in its maximum. Moreover, the condition $\xi /p\ll 1$,
with $\xi \lesssim 1$, holds for both, the low- and high-frequency regimes.
When $\xi \sim 1$ (so that $\omega \sim \tau _c^{-1}$), we get the
high-frequency regime $\hbar \omega \gg k_BT$, whereas for $\xi \ll 1$ even
the low-frequency regime is allowed. For the latter case we have to assure
that $0\ll \omega \ll \tau _c^{-1}$, not only to get rid of the divergence
arising from $\langle n\rangle _{\omega \text{ }}$, but also to hold the
assumption of highly excited oscillations of the CBR leading to the
Markovian approximation. Summarizing, under the conditions established above
we get the result ${\cal I}\sim 1+2\langle n\rangle _\omega $, which holds
for high- and low-frequency regime.
%%%%%%%%%%%%%%%%%%%%%%%%%%%%%%%%%%%%%%%%%%%%%%%%%%%%%%%%%%%%%%%%%%%%%%%%%%%%%%%%%%%%%%%%%%
%

\end{document}